\documentclass[journal]{vis/vgtc}                 
\ifpdf
  \pdfoutput=1\relax                   
  \usepackage{graphicx}                
  \DeclareGraphicsExtensions{.pdf,.png,.jpg,.jpeg} 
\else
  \usepackage{graphicx}                
  \DeclareGraphicsExtensions{.eps}     
\fi%

\graphicspath{{figures/}{pictures/}{images/}{./}} 

\usepackage{microtype}                 
\PassOptionsToPackage{warn}{textcomp}  
\usepackage{textcomp}                  
\usepackage{mathptmx}                  
\usepackage{times}                     
\usepackage{cite}                      
\usepackage{tabu}                      
\usepackage{booktabs}                  
\usepackage{enumitem}
\usepackage{listings}
\usepackage{color}
\usepackage{amsmath}
\usepackage{amssymb}
\usepackage{wrapfig}

\usepackage{tabularx}

\usepackage{makecell}
\usepackage[table]{xcolor}

\usepackage{xcolor}

\usepackage{framed,lipsum}

\usepackage[T1]{fontenc}
\usepackage[scaled=0.88]{beramono}    
\usepackage{listings}
\onlineid{1168}

\vgtccategory{Research}
\vgtcpapertype{Theoretical \& Empirical}

\vgtcinsertpkg

\title{A Grammar of Hypotheses for Visualization, Data, and Analysis}

\author{Ashley Suh, Ab Mosca, Eugene Wu, Remco Chang}
\authorfooter{
\item
Ashley Suh and Remco Chang are with Tufts University.  E-mail: \{ashley.suh, remco.chang\}@tufts.edu
\item
Ab Mosca is with Northeastern University. E-mail: a.mosca@northeastern.edu
\item
Eugene Wu is with Columbia University. E-mail: ewu@cs.columbia.edu.}

\teaser{
  \centering
  \includegraphics[width=0.95\textwidth]{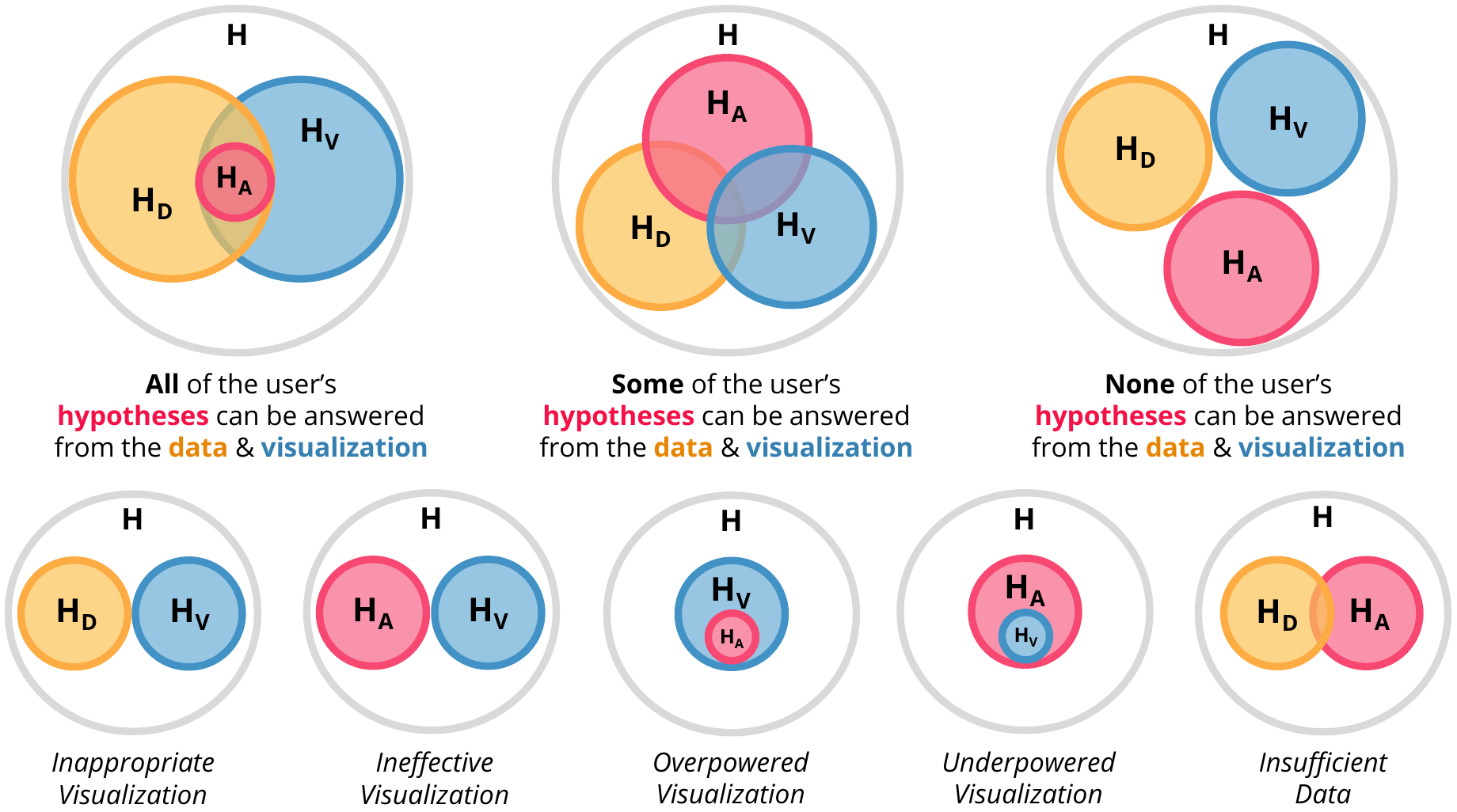} 
  \caption{The relationships between three \textit{hypothesis spaces}, formalized using our proposed \textit{hypothesis grammar}. A \textbf{visualization hypothesis space} (H$_V$) expresses the set of hypotheses a visualization can evaluate, a \textbf{data hypothesis space} (H$_D$) expresses the set of hypotheses a dataset can evaluate, and an \textbf{analysis hypothesis space} (H$_A$) expresses the set of hypotheses a user wants to evaluate through analysis. The intersection of these spaces can help us understand how the hypotheses a user would like to answer through analysis can either be fully answered by data and visualization, partially answered, or not answerable at all.}
  \label{fig:spaces}
}

\CCScatlist{
  \CCScatTwelve{Human-centered computing}{Visualization}{Empirical studies in visualization}{};
  \CCScatTwelve{Human-centered computing}{Visualization}{Visualization theory, concepts and paradigms}{}
}

\usepackage{amsthm}
\newtheoremstyle{exampstyle}
  {3pt} 
  {3pt} 
  {\itshape} 
  {} 
  {\bfseries} 
  {.} 
  {.5em} 
  {} 

\theoremstyle{exampstyle}

\usepackage{wrapfig}
\usepackage{amsmath}
\usepackage{amsfonts}
\usepackage{amssymb}
\usepackage{multirow}
\usepackage{enumitem}
\usepackage{listings}
\usepackage{booktabs} 
\usepackage{xspace}
\usepackage{alltt}
\usepackage{xcolor}
\usepackage{microtype}
\usepackage{mathtools}
\usepackage{subcaption}
\usepackage{balance}
\usepackage{tikz}
\usepackage[noend]{algpseudocode}
\usepackage{algorithmicx,algorithm}
\usepackage{comment}
\usepackage{microtype}
\usepackage{pgffor}
\usepackage{mwe}
\usepackage{todo}

\newtheorem{definition}{Definition}

\newlength{\listingindent}                
\lstdefinelanguage{goh}
{morekeywords={var,hyp,attr,const,expr,func,pred,op,except,
               expr1,expr2,var1,var2,attr1,attr2,const1,const2,const3,num1,num2,hyp1,hyp2},
sensitive=false,
morecomment=[l]{//},
morecomment=[s]{/*}{*/},
morestring=[b]",
}

\usepackage[noabbrev,capitalise]{cleveref}

\definecolor{light-gray}{gray}{0.95}
\definecolor{mid-gray}{gray}{0.85}
\definecolor{darkred}{rgb}{0.7,0.25,0.25}
\definecolor{darkgreen}{rgb}{0.15,0.55,0.15}
\definecolor{darkblue}{rgb}{0.1,0.1,0.5}
\definecolor{blue}{rgb}{0.19,0.58,1}
\definecolor{mred}{rgb}{.80,.12,.30}

\newcommand{\gray}[1]{\textcolor{mid-gray}{#1}}

\newcommand{\isnew}[1]{\textcolor{mred}{\textbf{#1}}}

\definecolor{grey}{rgb}{.5,.5,.5}
\definecolor{turquoise}{rgb}{.04,.79,.93}

\newif\ifnotes
\notestrue

\usepackage[normalem]{ulem}
\usepackage{soul}
\usepackage{hyperref}

\let\origcite\cite
\renewcommand{\cite}[1]{\ifnotes\mbox{\origcite{#1}}\else \origcite{#1}\fi}

\lstdefinestyle{SQLStyle}{
  language=SQL,
  showspaces=false,
  basicstyle=\ttfamily\scriptsize,
  commentstyle=\color{gray},
  mathescape=true,
  numbers=none,
  escapeinside={^}{^},
  captionpos=b,
  float=tp,
  floatplacement=tbp,
  belowskip=-0.05em,
   mathescape=false
}

\newlength{\leftbarwidth}
\newlength{\leftbarsep}
\colorlet{leftbarcolor}{gray}

\newcommand{\eat}[1]{}

\newcommand{\stitle}[1]{\vspace{2pt}\noindent\textbf{#1}}

\renewcommand{\manuscriptnotetxt}{}

\begin{document}

\abstract{%
We present a grammar for expressing hypotheses in visual data analysis to formalize the previously abstract notion of ``analysis tasks." Through the lens of our grammar, we lay the groundwork for how a user’s data analysis questions can be operationalized and automated as a set of hypotheses (a \textit{hypothesis space}). We demonstrate that our grammar-based approach for analysis tasks can provide a systematic method towards unifying three disparate spaces in visualization research: the hypotheses a dataset can express (a \textit{data hypothesis space}), the hypotheses a user would like to refine or verify through analysis (an \textit{analysis hypothesis space}), and the hypotheses a visualization design is capable of supporting (a \textit{visualization hypothesis space}). We illustrate how the formalization of these three spaces can inform future research in visualization evaluation, knowledge elicitation, analytic provenance, and visualization recommendation by using a shared language for hypotheses. Finally, we compare our proposed grammar-based approach with existing visual analysis models and discuss the potential of a new hypothesis-driven theory of visual analytics.
}
\maketitle


\section{Introduction}
\label{sec:introduction}
Visual analytics has long been hallmarked for its ability to support users through complex analysis tasks, such as training and debugging machine learning models\cite{sacha2019vis4ml}, hypothesis testing\cite{keim2006challenges}, knowledge generation\cite{sacha2014knowledge}, and decision-making\cite{dimara2021critical}. Despite this ubiquity, many contributed visual analysis systems are mechanistically agnostic to the user's analysis tasks, and ultimately their goals. The challenge of defining and operationalizing ``tasks'' in visual analysis has been an open problem for more than 40 years, with the first mention of the problem dating back to J. Mackinlay's proposed Presentation Tool (APT) in 1986\cite{mackinlay1986automating}.
Since then, numerous papers have attempted to address this challenge\cite{amar2005low, brehmer2013multi, miksch2014matter, laha2015classification, etemadpour2015user, sarikaya2018scatterplots, schulz2013design, isenberg2013systematic, sedlmair2012design, chang2010learning, dimara2018task}. 

Despite these efforts, the definition of ``task'' in visual analysis remains abstract. Originally used to describe data retrieval operations such as \textit{Lookup Value}, or \textit{Compare Values}, tasks have grown to include statistical and machine learning operations\cite{hu2019vizml} (e.g., identify clusters, discover outliers) and social interactions (e.g., \textit{Present Results} or \textit{Enjoy a Visualization}\cite{brehmer2013multi}). While this richness allows us to describe the plethora of ways in which tasks proliferate a user's interactions with visualization, these broad and diverse definitions of tasks make it increasingly difficult to operationalize them for visual analytics.

An operationalizable definition of ``tasks'' in visualization can have a profound impact on visualization research.
In addition to its application to automatic visualization recommendation\cite{mackinlay1986automating}, a composable definition of tasks can lead to potential breakthroughs in visualization design, evaluation, analysis provenance, natural language analytic tools, as well as knowledge and belief elicitation,. Conversely, without this definition, our community will continue to see a disconnect between empirical research on visualization tasks, and the practical implementation of that research in visual analytic tools.



In this paper, we propose a hypothesis grammar that builds a basis for operationalizing tasks in visual data analysis.
Our grammar is grounded in the concept of \textit{scientific hypotheses}, which offer a practical and well-defined starting point for concretizing the relationship between analysis tasks, data, and visualization. Through this formalism, we are able to describe and analyze three spaces (or sets) that can be expressed with our grammar: the hypotheses a dataset can evaluate (a \textit{data hypothesis space}), the hypotheses a user would like to evaluate through analysis (an \textit{analysis hypothesis space}), and the hypotheses a visualization design can evaluate (a \textit{visualization hypothesis space}). 
{The development and exploration} of the hypothesis grammar takes a step towards making tasks operational, allowing us to unify three areas (data, analysis tasks, visual encodings) that are often researched in isolation.
The unification of these spaces
enables a new methodology for the visualization community to reason about concepts that had not been possible or easy to do previously, including a hypothesis-driven approach to analysis.

For example, during analysis, a user could express their questions using our grammar, then examine whether these questions are answerable with the available data and the visualization tool. 
Concepts such as ``\textit{exploration},'' ``\textit{insight},'' and ``\textit{knowledge verification}'' can be captured through the relationship and intersection of these hypothesis spaces (e.g., gaining new insights might ``shrink'' the analysis hypothesis space by ``removing'' hypotheses that are irrelevant to the user). 
Visualizations can be formalized, not by their data encodings, but in terms of the hypotheses that the visualizations can be used to verify, thereby enabling the possibility of a novel 
hypothesis-based theory of visualization.


In the following sections, we elaborate on the merits of modeling analysis tasks using a grammar-based approach for scientific hypotheses. 
We then introduce our proposed grammar, and demonstrate how visualization experts can use it to reason about the 
interconnection of data, analysis tasks, and visual encodings.
We discuss the formalization of three hypothesis spaces generated with our grammar, and how they can spur new directions in visualization research. Finally, we provide a demonstration of our grammar through an analysis case study, and conclude with a direct comparison to existing work in this area.


\smallskip\noindent To summarize, we make the following contributions in this paper: 

\begin{itemize}[topsep=1pt, partopsep=1pt,itemsep=1pt,parsep=1pt]
    \item We introduce a Grammar of Hypotheses to operationalize analysis tasks with respect to a user's questions, data, and visualization.
    \item We use the grammar to formalize three ``hypothesis spaces'': an \textit{analysis hypothesis space}, a \textit{data hypothesis space}, and a \textit{visualization hypothesis space}.
    \item We present case studies to demonstrate how the unification of these three spaces can inform future visualization research.
    \item We provide a comparison to existing visual analysis theories to illustrate how our hypothesis grammar augments current practices, suggesting a new exploratory hypothesis analysis methodology.
\end{itemize}

\section{Background \& Motivation}
\label{sec:hypothesis}
The concept of a ``hypothesis'' has varying implications depending on the context and community in which it is used. 
In this section, we provide a brief background on the use of hypotheses and grammars in the context of the visualization and science communities. From this background, we motivate why a grammar-based approach -- specifically, a grammar of hypotheses -- is an advantageous method to begin operationalizing visual analysis tasks.

\subsection{Hypotheses} 

\subsubsection{Scientific Hypotheses}
\label{sec:background-scientific-hypotheses}
In the science community, the precise criteria for a scientific hypothesis has been well-studied\cite{quinn1975teaching, sunal2002social} and is rigorously defined as a proposition (or set of propositions) proposed as a tentative explanation for an observed situation or phenomenon\cite{gauch2003scientific, cohen2011introduction}. A \textit{valid} scientific hypothesis must be both \textit{testable} and \textit{falsifiable}\cite{popper2005logic}. 
Researchers 
have proposed a number of requirements for evaluating the quality of a scientific hypothesis\cite{van2007issues, mulder2010finding}. For example, work by Quinn and George\cite{quinn1975teaching} states that a ``good'' hypothesis must have the following properties: 

\begin{itemize}[topsep=1pt, partopsep=1pt,itemsep=1pt,parsep=1pt]
    \item it makes sense; 
    \item it is empirical, a (partial) scientific relation;
    \item it is adequate, a scientific relation between at least two variables; 
    \item it is precise—a qualified and/or quantified relation; 
    \item it states a test, an explicit statement of a test. 
\end{itemize}


Notably, research in this area
distinguishes  between \textbf{scientific hypotheses} and \textbf{statistical hypotheses}\cite{toledo2011developing, jeong2006definition}. Scholars have argued that the two are distinct due to statistical hypotheses' use of confidence thresholds (e.g., p-values) and the reliance on null and alternate hypotheses -- both of which are inappropriate for scientific hypotheses\cite{bolles1962difference}.
In this work, we base our grammar on the definition and criteria for scientific hypotheses. This provides (1) a formal definition 
on which a grammar can be based, and (2) structure for thinking about hypothesis generation and refinement in visual data analysis. 


\subsubsection{Significance of Hypotheses in Visual Analytics} 

A central goal of visual analytics is supporting users through forming, refining, and validating hypotheses\cite{keim2006challenges, keim:2010:mastering}. In the \textit{Sensemaking} process\cite{pirolli2005sensemaking}, the iterative refinement of hypotheses is an integral step before the presentation of analysis findings. In the \textit{Knowledge Generation Model}\cite{sacha2014knowledge}, a hypothesis is generated from knowledge and insight, and is the driving motivation behind a user's (inter)actions with a visual analytics system. Pike et al.\cite{pike2009science} write, ``\textit{[in] constructing hypotheses or `abductions', the analyst is engaged in exploration of the data space and the formation of mental models to explain observations.}'' 

These theoretical models reflect how the visual analytics community considers the concept of \textit{hypotheses} integral to effective design. However, while many visual analytic models discuss the importance of supporting hypothesis generation and verification, there are few operational means of both supporting and assessing hypotheses in visual data analysis. 
The grammar proposed in this paper allows us to combine the ethos of visual analytics hypotheses with the rigor and precision of scientific hypotheses, resulting in a more concrete way to model and reason about sensemaking and visual analysis. We discuss the relationship of our grammar with existing theories in Section~\ref{sec:comparison}.

\subsection{Grammars} 
A grammar can disambiguate the subjective view of a user's task, allowing for new designs of visualization systems to reason and operate over both data and the user's analysis goals. 
We discuss how grammars have been used with success in the science and visualization communities.

\subsubsection{Use of Grammars in Science} 

Grammars for representing the syntax of hypotheses have been used previously for science education\cite{gijlers2005relation, mulder2010finding}.
A large focus is on building tools (on top of a grammar) to help students build expressions from variables (i.e. data attributes) and relations (e.g., `increases' or `positively correlated with') to create a hypothesis that meets the criteria discussed in Section~\ref{sec:background-scientific-hypotheses} (e.g., if A increases, then B increases). 

For example, Kroeze et al.\cite{kroeze2019automated} 
use a context-free grammar (CFG) to automatically parse and give (semantic) feedback on a student's hypotheses by using the parsing tree.
A subset of this CFG (in the domain context of electrical circuits) is:
\begin{lstlisting}
        HYPOTHESIS -> if ACTION then ACTION
        ACTION     -> VAR MODIFIER
        VAR        -> voltage | brightness
        MODIFIER   -> increases | decreases
\end{lstlisting}

An example hypothesis parsable from this CFG includes ``\textit{if voltage increases, then brightness increases}.'' 
In the development of our hypothesis grammar, we drew inspiration from many works in science education\cite{de2006technological, van2007issues}. We present our own version of a hypothesis grammar, but applied more broadly to analysis tasks within the context of visualization and visual analysis, in Section~\ref{sec:grammar}.

\subsubsection{Grammars in Visualization \& Visual Analytics} 
A core theory in visualization today is the concept of a graphical grammar, which decomposes a data visualization into orthogonal components that are easy to learn and automatically analyze, implement, and reason about.  
The use of grammars in visualization was popularized by the development of the Grammar of Graphics by Leland Wilkinson\cite{wilkinson2012grammar} in 1999.
The Grammar of Graphics provides a set of rules such that a visualization is constructed by mapping data attributes to visual properties, e.g., position, size, color, and shape. The Grammar of Graphics has been used as the basis for several popular visualization tools, such as R's ggplot2\cite{wickham2011ggplot2}, and has inspired the development of many other visualization grammars\cite{li2020gotree, park2017atom, wongsuphasawat2020encodable, satyanarayan2014declarative, mcnutt1912no}.

The use of grammars has enabled the development of powerful and flexible visualization tools that can be easily customized to meet the needs of diverse users and applications. 
Grammars have also provided a new means to reason about visualization design and theory. By building on top of the Grammar of Graphics, Vega-Lite\cite{satyanarayan2016vega} elevated interaction as a first-class citizen to design, allowing the specification of visualizations and dashboards to include interactivity. Systems like Voyager 2\cite{2017-voyager2} and Draco\cite{moritz2018formalizing} similarly leverage the flexibility and robustness of grammars to enhance visualization recommendation.

Grammars have helped the visualization research community take great strides in supporting users through the construction of custom visualizations with concise, intuitive, structured, and consistent languages. They have also facilitated new theories to reason over visual encodings and interaction techniques. However, we observe that visualization grammars used today are still agnostic to a user's analysis task -- despite the fundamental principle that supporting a user's task is essential to good visualization design\cite{sedlmair2012design}. 
Moreover, understanding which visualizations excel at answering questions beyond simple analytic tasks\cite{amar2005low, saket2018task}, or evaluating hypotheses (and to what capacity) is difficult to reason about.
In Section~\ref{sec:implications}, we illustrate how a hypothesis grammar can help bridge these gaps.

\subsection{Why a Grammar of Hypotheses?}
\label{sec:background-why-ours}
There is a plethora of domain-specific languages related to data manipulation (and data visualization), ranging from database query languages such as SQL\cite{chamberlin1974sequel} and DataLog\cite{abiteboul1995foundations}, to recent business intelligence languages like Malloy~\cite{malloy}.
These languages are expressive and practically useful. 
However, from the perspective of visualization research, their expressivity and generalizability make them difficult to critically think through the relationships between data, tasks, and visualization.

Our proposed hypothesis grammar is designed with the opposite goals in mind.
We do not intend for the grammar to be implemented as a practical programming language at this moment. Instead, it is deliberately made narrow to reason about visualization, visual data analysis, and operational tasks. 
We consider the simplicity of the grammar a desirable property, as it provides a \textit{minimum viable product} to illustrate how we can resolve a subset of complex problems that visualization researchers have been grappling with for decades\cite{mackinlay1986automating}.

It is relevant to note that the design of our proposed hypothesis grammar is inspired by SQL.
In fact, the hypothesis grammar can be seen as a very limited subset of SQL where every statement in the grammar can be trivially converted to and evaluated with SQL.
As such, all examples shown in this paper can be executed by a relational database system with minimum modification. We discuss these implications, and considerations for other languages that could be used, in Section~\ref{sec:discussion}.
\begin{table*}[t!]
    \centering
    \renewcommand{\arraystretch}{1.3}
    \centering
    \resizebox{.9\linewidth}{!}{%
    
\definecolor{palesilver}{rgb}{0.9, 0.9, 0.9}



\sffamily
\begin{tabular}{p{0.05\linewidth} p{0.4\linewidth} p{0.55\linewidth}}
\toprule

\textbf{Term} & 
\textbf{Production Rule} & 
\textbf{Semantic Definition} \\

\hline
\rowcolor{palesilver}

\verb|hyp| & \texttt{expr op expr ("["pred"]")? ("\&" hyp)?} & A statement, which evaluates to either \verb|true| or \verb|false|, describing the relationship between at least one expression and one variable \\

\verb|expr| & \texttt{func "(" (expr ("," expr)? )? ")" | var} & A nullary, unary, binary expression, or variable. \\

\rowcolor{palesilver}

\verb|var| & \texttt{attr ("[" pred "]")? | const} & An attribute reference or a constant \\

\verb|pred| & \texttt{var op const (\& pred)?} & A condition expression that evaluates to \verb|true| or \verb|false| for a given \verb|var|\\

\rowcolor{palesilver}

\verb|func| & \verb|str| & The name of a function with explicit input and output types from a predefined library or table.   \\

\verb|op| & \texttt{= | < | > | ...} & A boolean operation. \\

\rowcolor{palesilver}
\verb|attr| & \verb|str| & An attribute reference.  Grounded to attributes in dataset. \\

\verb|const| & \texttt{str | number | .. | array } & Constants are specified as part of the hypothesis, and may be any scalar, or array  that isn't derived from the database.  \\

\hline
\end{tabular} 
}
    \caption{Breaking down each term in our proposed hypothesis grammar (referred as the \textbf{Base Grammar}), as described in Section~\ref{sec:grammar}.
    }
    \label{tab:grammar_definitions}
\end{table*}

\section{Hypothesis Grammar}
\label{sec:grammar}

Before discussing the implications of a grammar-based approach for tasks, 
we first need to establish a baseline with our hypothesis grammar. We start with an illustrative example, showing how the hypotheses discussed in Section~\ref{sec:hypothesis} can be formed from a simple tabular dataset and toy grammar. We then walk through a generalizable hypothesis grammar that is capable of expressing hypotheses from tabular datasets. 
We present this grammar in the most simple representations as possible, and leave nuances and complexities for supplemental material.

\subsection{Illustrative Hypothesis Grammar}
Before introducing the full hypothesis grammar, we begin with small, illustrative examples. The complete grammar is presented in Section~\ref{sec:grammar-complete}. 

\subsubsection{Definition}
\label{sec:grammar:illustrative:definition}
Our simple grammar is based on the definition of a hypothesis, specifically, that a hypothesis is ``\textit{a qualifiable relation between at least two variables.}''
The goal is to design a grammar that can express relations such as: \lstinline{average(sales)>average(cost)} or \lstinline{average(cost)>10}. 
Based on this definition, we define a hypothesis as:

\begin{lstlisting}
     hyp    :- expr op expr [pred]
     expr   :- const | attr | func(expr) | func(expr, expr) 
     op     :- > | < | = | ...
     func   :- AVG | MAX | MIN | ...
     pred   :- attr op const
     const  :- number
\end{lstlisting}
In this grammar, a hypothesis (\lstinline{hyp}) is defined with expressions (\lstinline{expr}).
An expression can be a data attribute (\lstinline{attr}) such as \lstinline{sales}, a constant (in this case, a \lstinline{number}), or a function (\lstinline{func}) over another expression.
Data attributes (\lstinline{attr}) are currently unresolved in this grammar, which we will resolve later in Section~\ref{sec:spaces}.

Since the evaluation of a hypothesis results in a binary true or false, the operator (\lstinline{op}) is limited to binary relations (such as \lstinline{>}, \lstinline{<}, \lstinline{=}, etc.).
The list of functions for a hypothesis grammar needs to be preregistered, similar to registering a user-defined function in a SQL database.
For simplicity, we assume that the list of functions includes the typical aggregation (such as \lstinline{AVG}, \lstinline{SUM}, \lstinline{MIN}, etc.) and analytic functions (such as \lstinline{CORR} for correlation, \lstinline{STDDEV} for standard deviation, etc.) that are commonly supported by SQL databases.

Lastly, we introduce the notion of a predicate (\lstinline{pred}), which functions similarly to a \lstinline{WHERE} clause in SQL queries to filter data. 
For example, a predicate can express \lstinline{[year=2023]} to filter data by the year 2023.

\subsubsection{Examples}
We provide examples of the use of this grammar to express hypotheses:
\begin{itemize} [topsep=3pt, itemsep=0em]
    \item{``\textit{Is my company profitable?}'' can translate to the hypothesis ``the total amount of sales is higher than cost and expenditure.'' This is formulated as:\\ \lstinline{SUM(sales) > ADD(SUM(cost), SUM(expenditure))}.
    }
    
    \item ``\textit{If the temperature of an object goes up, its volume will go up}'' is a scientific hypothesis.  It can be expressed as ``there is a positive correlation between temperature and volume.'' Let \lstinline{CORR} compute the correlation between two variables.  We can reformulate as:\\ \lstinline{CORR(temperature, volume)>0.9}. 
    
    \item ``\textit{Car horsepower (hp) since 2000 follows a normal distribution.}'' Let \lstinline{KS-NORMAL} be the Kolmogorov Smirnov test with alpha value 0.05 to evaluate if a variable follows a normal distribution. This hypothesis can be formulated as:\\  \lstinline{KS-NORMAL(hp)>0.05 [year>2000]} 
\end{itemize}
Beyond these examples, we also demonstrate in Section~\ref{sec:demonstration} how the hypothesis grammar can be applied to analyze the VAST 2017 Challenge Mini-Challenge 1 (the Boonsong Kekagul Nature Reserve challenge).

\subsection{Complete Hypothesis Grammar}
\label{sec:grammar-complete}
Our complete hypothesis grammar (in PEG notation\cite{ford2004parsing}) is shown in Table~\ref{tab:grammar_definitions}. Other notations could be used, we discuss in Section~\ref{sec:discussion}. The complete grammar (which we refer to as our \textbf{base grammar}) is similar to the illustrative grammar, but with four minor changes:

\begin{itemize} [topsep=3pt, itemsep=0em]
\item (\lstinline{hyp & hyp}): A hypothesis is now recursively defined such that it can contain multiple sub-hypothesis. In other words, a hypothesis can now be written as \lstinline{hyp :- $H_1$ & $H_2$ & $H_3$ & ...} where each of the hypothesis $H_i$ is separately defined as \lstinline{expr op expr}.  
\item (\lstinline{var}): Each expression (\lstinline{expr}) in a hypothesis (\lstinline{hyp}) can have its own predicate. This allows independent filtering for the two expressions. For example, \lstinline{AVG(price)[year<2000] < AVG(price)[year>=2000]} is now possible. This now introduces a new symbol called \lstinline{var} where \lstinline{var:- attr [pred]}.  
\item (\lstinline{pred & pred}): Similarly, a predicate can be recursively defined, and can be written as \lstinline{pred :- $P_1$ & $P_2$ & $P_3$ & ...} where each of the predicates $P_i$ is separately defined as \lstinline{var op const}.  
\item (\lstinline{func(expr, expr,...)}): The number of parameters to a function is now unlimited. This allows for the following expression \lstinline{func(expr$_1$, expr$_2$, expr$_3$,...)} where each \lstinline{expr$_i$} is a parameter of to the function.
\end{itemize}

We note that the immediate, widespread implementation of the grammar was not the primary focus when we designed our hypothesis grammar.
Instead, we use the hypothesis grammar as a way to illustrate how a grammar based on hypotheses can be used to reason about properties and concepts in visualization research.
In the section below, we describe how the grammar can formalize and unify the concepts of data, analysis tasks, and visualization.

\section{Hypothesis Spaces}
\label{sec:spaces}

We first observe that the hypothesis grammar can be used as a set of production rules to generate valid hypotheses -- that is, statements that are acceptable (parsable) by the hypothesis grammar.
Conversely, we can consider all possible hypothesis statements that are acceptable by the hypothesis grammar as a set, which we refer to as a \textbf{hypothesis space}.
We use this notion to describe the hypotheses spaces of a dataset (a \textit{data hypothesis space}), a visualization (a \textit{visualization hypothesis space}), and an analyst's analysis tasks (an \textit{analysis hypothesis space}). 

We illustrate each of these spaces using PEG notation, while omitting the quotations for simplicity. Table~\ref{tab:grammar_definitions} shows all full marks.




\subsection{Full Hypothesis Space (\texorpdfstring{$H$}{H})}
The full (infinite) hypothesis space is denoted as $H$ and represents \textit{all} possible hypotheses that can be generated. 
Formally, it is defined as all statements that can be parsed by the base grammar (see Table~\ref{tab:grammar_definitions}). 

\subsection{Data Hypothesis Space (\texorpdfstring{$H_D$}{HD})}
\label{sec:spaces:data}
\begin{wrapfigure}{l}{.08\columnwidth}
    \centering
    \vspace{-15pt}
    \includegraphics[width=.11\columnwidth]{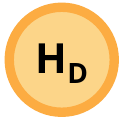} 
    \vspace{-20pt}
\end{wrapfigure}
A data hypothesis space is denoted as $H_D$. 
The space is defined by `grounding' the base grammar to a dataset $D$.
Informally, we say that $H_D$ describes all hypotheses that can be evaluated or verified using the dataset $D$, therefore, $H_D$ is a strict subset of $H$.

\begin{wraptable}{r}{2cm}
\vspace{-10pt}
{\small
\begin{tabular}{ c c }
    \centering
     \texttt{cost} & \texttt{price}\\ 
     \midrule
     \texttt{750} & \texttt{350} \\ 
     \texttt{1000} & \texttt{425} \\ 
     \texttt{850} & \texttt{500} \\ 
     \texttt{600} & \texttt{250} \\[3pt] 
\end{tabular}
\vspace{-25pt}
}
\end{wraptable}
To illustrate this, consider a simple dataset (shown to the right) with two numerical attributes, \lstinline{cost} and \lstinline{price}. 
As shown below, we can extend the base grammar by grounding it to this data. 
The highlighted text in red represents changes to the original grammar in Section~\ref{sec:grammar-complete}.

\begin{lstlisting}    
     hyp    :- expr op expr ([pred]) (& hyp)?
     expr   :- func ((expr (, expr)?)? ) | var
     var    :- attr ([pred])? | const
     pred   :- var op const (& pred)?
     func   :- AVG | MAX | MIN | ...
     op     :- > | < | = | ...
     const  :- number 
     ^\isnew{attr}^   :- ^\isnew{cost}^ | ^\isnew{price}^
\end{lstlisting}

Although the change seems minimal, the implication of grounding the data to the grammar is significant.
Assuming a finite set of operators and functions, this grammar describes \textit{all possible hypotheses} that we can now pose with this simple two-attribute dataset.

We note that the number of hypotheses is not finite due to the recursive expansions of the \lstinline{expr} rule and the infinite possible resolutions of \lstinline{const}.
The grammar can be made finite by applying restrictions on the depth of the recursion and binning the possible values.

Given this grounded grammar, we can now define a set of verifiable hypotheses (acceptable statements) based on whether the grammar and dataset are able to parse each statement. 

\subsection{Analysis Hypothesis Space (\texorpdfstring{$H_A$}{HA})}
\label{sec:analyst-hypothesis-space}
\begin{wrapfigure}{l}{.08\columnwidth}
    \centering
    \vspace{-15pt}
    \includegraphics[width=.11\columnwidth]{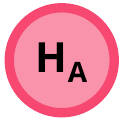} 
    \vspace{-20pt}
\end{wrapfigure}
An analysis hypothesis space is denoted as $H_A$.
This space represents the set of analysis questions (expressed as hypotheses) that a user wants to evaluate, independent of the data that is available to them, or the consideration of what visualization to use.

To illustrate $H_A$, consider the previous grammar used for $H_D$.
If a user thinks that \lstinline{interest-rate}, the \lstinline{year}, and \lstinline{stock-price} might be correlated with \lstinline{cost}, she might add them along with \lstinline|CORR| to the previous grammar as follows: 
\begin{lstlisting}   
     hyp   :- expr op expr ([pred]) (& hyp)?
     expr  :- func ((expr (, expr)?)?) | var
     var   :- attr ([pred])? | const
     pred  :- attr op const (& pred)?
     func  :- AVG | MAX | MIN | ^\isnew{CORR}^ | ...
     op    :- > | < | = | ...
     const :- number
     attr  := cost | ^\isnew{interest-rate}^ | ^\isnew{year}^ | ^\isnew{stock-price}^
\end{lstlisting}

This grammar represents, in the form of hypotheses, the analysis questions that a user might have. 
It is important to note that the attributes \lstinline{interest-rate}, \lstinline{year}, and \lstinline{stock-prices} are not in the dataset, and simply represent attributes that the user is interested in. 
Similarly, the attribute \lstinline{price} has been removed from the grammar because it is not an attribute that the user cares about in this analysis.
In this regard, the analysis hypothesis space can be created independently of the data.

In Section~\ref{sec:demonstration}, we use the 2017 VAST Challenge to demonstrate how realistic analysis tasks can be formalized as $H_A$.

\subsection{Visualization Hypothesis Space (\texorpdfstring{$H_V$}{HV})}
\label{sec:vis-space}
\begin{wrapfigure}{l}{.08\columnwidth}
    \centering
    \vspace{-15pt}
    \includegraphics[width=.11\columnwidth]{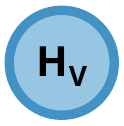} 
    \vspace{-20pt}
\end{wrapfigure}
The visualization hypothesis space, denoted $H_V$, represents all hypotheses that a visualization $V$ is able to evaluate. 
Similar to how $H_A$ can be created without the consideration of the data, here we consider $H_V$ independently of any dataset as well.  
%
To illustrate the general concept of $H_V$, we provide examples of visualization hypothesis spaces below. 

\stitle{Hypothesis Space of a Barchart with 3 Bars:} 
Consider $V_b$, a static barchart graphic that shows three bars (shown left). 
\begin{wrapfigure}{l}{.35\columnwidth}
    \centering
    \vspace{-8pt}
    \includegraphics[width=.37\columnwidth]{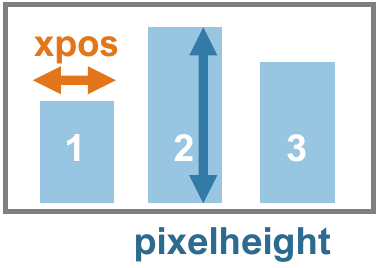} 
    \vspace{-22pt}
    \label{fig:barchart}
\end{wrapfigure}
\noindent
When $V_b$ is rendered as an image, the image has specific property values for the canvas (width, height, border size, spacing, etc.) and contains three bars that have positional attributes, such as their position along the x-axis in pixels (\lstinline{xpos}) and their height in pixels (\lstinline{pixelheight}).  
For now, let's only focus on these two attributes.
\hfill
$V_b$ can be used to evaluate a few different hypotheses. For example, whether bar$_{1}$ is taller than bar$_{2}$ (\lstinline{pixelheight[id=1] > pixelheight[id=2]}) or if the three bars show an increasing order in height (\lstinline{fit(pixelheight, xpos) > const}). 
Given the limited hypotheses expressible by $V_b$, the hypothesis space should be relatively small and can be expressed as: 
\begin{lstlisting}
     hyp   :- hyp1 | hyp2 
     hyp1  :- pixelheight[id= num1] op 
              pixelheight[id= num2 & num1 ^$\neq$^ num2]
     hyp2  :- fit(pixelheight, xpos) > const
     attr  :- pixelheight | xpos
     op    :- = | < | > | ...
     num1  :- number             num2 :- number
     const :- number
\end{lstlisting}

This grammar (and the hypothesis space) is specific to this instance of $V_b$, and will change if its design specification or dataset changes.  For example, if $V_b$ now renders 100 bars, evaluating if the barchart exhibits a normal distribution may be a valid hypothesis. 
Conversely, it may no longer be valid to compare the heights of bar$_5$ with bar$_{95}$ because perceptually this becomes much harder to do, especially if the width of the visualization is large\cite{talbot2014four}.

\stitle{Visualization Hypothesis Space and Graphical Perception:} 
Finding the limitations of human perception and cognition with different visual encodings is still an active research area in the visualization community.
There is, at this time, no definitive mapping between all visualization designs and their task effectiveness.
However, research has begun to propose equivalent design spaces (e.g.,\cite{harrison2014ranking, sarikaya2018scatterplots, pandey2016towards, yang2018correlation, correll2017regression}). 
Recent work such as Saket et al.\cite{saket2018task}, Kim et al.\cite{kim2018assessing}, and Mortiz et al.\cite{moritz2018formalizing} have started to empirically evaluate and model the effectiveness of visualizations for analytics tasks. 
Our grammar for describing the hypothesis space of a visualization currently cannot account for these perceptual models in a generalized way.
However, we will further discuss the implications of quantifying and evaluating visualization hypothesis spaces in Section~\ref{sec:discussion}.

\subsection{Definitions of Hypothesis Spaces}
To formalize the three hypothesis spaces and apply them to reason about the implications of the spaces, we first define some operations and properties of the hypothesis spaces: 

\begin{definition}[A Hypothesis]
A hypothesis statement ($h$) is said to be in a hypothesis space $H_1$ (i.e., $h \in H_1$) if $h$ can be parsed by the grammar associated with the hypothesis space. 
\label{def:statement}
\end{definition} 

\begin{definition}[Size of a Hypothesis Space]
The size of a hypothesis space $H_1$ is denoted as $|H_1|$, representing the number of statements that can be parsed by the grammar associated with the hypothesis space.  
\label{def:size}
\end{definition} 

\begin{definition}[Containment]
Given two hypothesis spaces $H_1$ and $H_2$, 
if $H_1\subset H_2$ then $|H_1| < |H_2|$.
Further, any solution that can be used to answer hypotheses in $H_2$ can be used to answer $H_1$.
\label{def:containtment}
\end{definition} 

\begin{definition}[Intersection]
Given two hypothesis spaces $H_1$ and $H_2$, their intersection $H_1 \cap H_2$ represents hypotheses that can be parsed by both the grammar of $H_1$ and the grammar for $H_2$. $H_1\not\cap~H_2$ denotes that there is no intersection between the two spaces (i.e., $|H_1 \cap H_2| = 0$). 
\label{def:intersection}
\end{definition} 
\vspace{-10pt}
\begin{definition}[Grounding]
Grounding a hypothesis space $H_1$ to a dataset $D$ is defined as finding the intersection between $H_1$ and $H_D$, where $H_D$ is the hypothesis space of $D$.
For example, grounding a user's analysis hypothesis space $H_A$ to $D$ results in a new hypothesis space $H_{A'}$ such that $H_{A'} = H_A \cap H_D$.
\label{def:grounding}
\end{definition}

\section{Implications of the Hypothesis Grammar}
\label{sec:implications}

In the visualization community, grammars have facilitated the formalisms for expressing visual encodings\cite{wilkinson2012grammar}, user interactions\cite{satyanarayan2016vega}, and visualization recommendations\cite{2017-voyager2, moritz2018formalizing}. 
We believe our proposed grammar can similarly help researchers reason about design, analysis, and tasks in terms of the hypotheses they support. Below, we discuss the implications of the relationship between the three hypothesis spaces, how they can be used to model a visual analysis process, and how the grammar can facilitate the formalization of analysis tasks.

\subsection{Unifying the Three Spaces}
\label{sec:implications-relationships}
Our proposed hypothesis grammar allows us to express data, visualization, and analysis using \textbf{the same language}. 
By doing so, we can consider the relationships produced by the unification of the hypothesis spaces $H_D$, $H_V$, and $H_A$ -- as well as their implications to the roles of data, visualization, and users in visual analytics.
Figure~\ref{fig:spaces} provides an illustration of some interesting relationships between these spaces: 

\medbreak

\begin{wrapfigure}{l}{.15\columnwidth}
    \centering
    \vspace{-15pt}
    \includegraphics[width=.17\columnwidth]{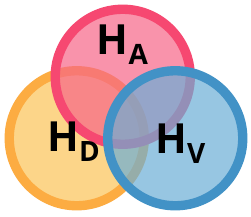} 
    \vspace{-22pt}
\end{wrapfigure}
\stitle{Successful Visual Analytics Solutions}. 
A visual analytics solution can be considered successful if the intersection between $H_D$, $H_V$, and $H_A$ is not null (i.e., $|H_D \cap H_V \cap H_A| > 0$). 
An ideal solution is when a user's questions can be fully answered by the data and the visualization tool such that $H_A \subset H_D$,  $H_A \subset H_V$, and $|H_D \cap H_V| > 0$.

\medbreak
\begin{wrapfigure}{l}{.15\columnwidth}
    \centering
    \vspace{-15pt}
    \includegraphics[width=.17\columnwidth]{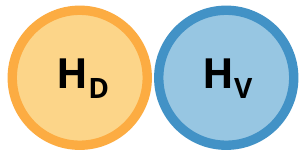} 
    \vspace{-20pt}
\end{wrapfigure}
\stitle{Inappropriate Visualization}.
The use of an \textit{inappropriate} visualization occurs when $H_V\not\cap~H_D$ (or $|H_V \cap H_D|$ is very small) where a visualization or visual encoding cannot support the hypotheses about the data. 

\medbreak
\begin{wrapfigure}{l}{.15\columnwidth}
    \centering
    \vspace{-15pt}
    \includegraphics[width=.17\columnwidth]{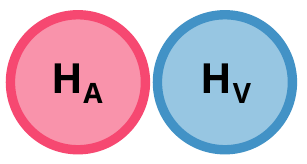} 
    \vspace{-20pt}
\end{wrapfigure}
\stitle{Ineffective Visualization}.
The use of an \textit{ineffective} visualization occurs when $H_V \not\cap~H_A$ (or $|H_V \cap H_A|$ is very small), that is, a visualization or visual tool cannot support a user's analysis questions. 

\medbreak
\begin{wrapfigure}{l}{.15\columnwidth}
    \centering
    \vspace{-15pt}
    \includegraphics[width=.17\columnwidth]{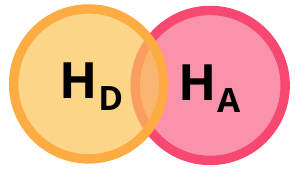} 
    \vspace{-20pt}
\end{wrapfigure}
\stitle{Insufficient Data}.
In the case where $H_A \not\subset H_D$, not all of a user's analysis questions can be answered with the given data. 
This might require acquiring additional data, or performing new data transformations.


\noindent 
\textbf{ } 

\begin{wrapfigure}{l}{0.075\textwidth}
    \centering
    \vspace{-16pt}
    \includegraphics[width=.12\columnwidth]{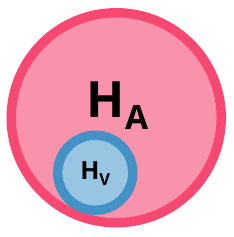} 
    \vspace{-15pt}
\end{wrapfigure}

\noindent 
\textbf{Under-Powered Visualization.} If $H_V \subset H_A$ but $|H_A|$ is significantly larger than $|H_V|$ (i.e., $|H_A| \gg |H_V|$), the visualization is too limited for the user's analysis needs, and a more robust visualization solution or tool is necessary. 



\noindent 
\textbf{ } 

\begin{wrapfigure}{l}{0.075\textwidth}
    \centering
    \vspace{-16pt}
    \includegraphics[width=.12\columnwidth]{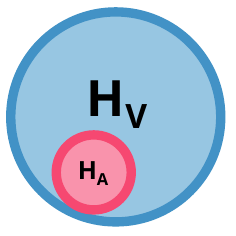} 
    \vspace{-15pt}
\end{wrapfigure}

\noindent 
\textbf{Over-Powered Visualization.} Conversely, if $H_A \subset H_V$ but $|H_V|$ is significantly larger than $|H_A|$ (i.e., $|H_V| \gg |H_A|$), the visualization is likely too complex for the user's needs, or can encode broad data and questions (e.g., dashboards). 

\hfill

\subsection{Partial Grammar Specification to Refine Analysis}
\label{sec:demonstration-partial}

In Section~\ref{sec:spaces}, we demonstrated how a grammar can be used to represent a user's analysis tasks, forming an analysis hypothesis space ($H_A$). Our grammar affords us the benefit of using non-terminals that can be resolved, e.g., with unification\cite{shieber2003introduction} (similar to the concept of \textit{wildcards}\cite{wongsuphasawat2016towards}). As a result, users can \textit{partially specify} their analysis goals or tasks through the use of our grammar. Visualization tools like Voyager 2\cite{2017-voyager2} have shown the success of using wildcards with a grammar to aid in visualization design and recommendation. We posit that similar methodologies with our hypothesis grammar can be used for analysis task and hypothesis refinement. 



Consider the following grammar (based on Section~\ref{sec:analyst-hypothesis-space}'s example):

\begin{lstlisting}   
     hyp    :- CORR(cost, attr) op const1 [price > const2] 
     attr   :- interest-rate | year | stock-price
     op     :- = | < | > | ...
     const1 :- number      const2 := number
\end{lstlisting}

In this example, the user has a narrow analysis task to find out what other attributes (\lstinline{attr}) might correlate with the attribute \lstinline{cost}. 
The user is only interested in items that have a price above some threshold; however, at this point the user does not have a definitive sense of what that threshold is (\lstinline{const2}).

The above grammar forms a hypothesis space that is smaller than the original one shown in Section~\ref{sec:analyst-hypothesis-space}.
We can consider the process of adding specificity to the analysis hypothesis grammar (thereby reducing the size of it) as an user's \textbf{refinement} of their analysis tasks towards resolving a goal. When the size of the analysis hypothesis space reaches one (or a similarly small number), the user should then be able to evaluate the hypotheses analytically\cite{keim2006challenges}.
Until then, 
the user interacts with the visual analysis system to explore $H_V$, grounded to their data.
During this process, the user is exposed to analysis tasks (in our formalism: the hypotheses supported by their visualization and data) of which they may not have been previously aware. In this way, our grammar is able to model \textit{vagueness} in analysis goals (e.g. the unknown \lstinline{const2}) and refinement as that vagueness is replaced with new knowledge (e.g. setting \lstinline{const2} to a reasonable threshold).    

\subsubsection{Beyond Refinement to Incorporate Insight}
In the above illustration, we focused on how partial grammar specifications can be used to clarify $H_A$. As the user obtains a better understanding of their analysis goals (e.g., resolving \lstinline{const2} to a real value, from the previous example), we show that $H_A$ similarly becomes more refined. This formulation follows similar analysis models\cite{tukey1977exploratory, pirolli2005sensemaking, keim2006challenges, sacha2014knowledge} that describe how high-level analysis goals are refined through \textit{exploration} until confirmatory analyses can occur. 

However, in practice, conducting analysis is rarely a linear process in which $H_A$ begins at one state (and thus one size) and continuously shrinks until $H_A$ becomes extremely small. 
Instead, it is more likely that $H_A$ starts at an initial state, represented as a grammar referred to as $H_{A_0}$, and moves across analysis states ($H_{A_0}$, $H_{A_1}$, ..., $H_{A_i}$) with corresponding hypotheses spaces that ``move,'' expand, and shrink in size at any point. This might happen because the user accesses new data ($H_D$ grows), or because the user gains new insights into the data and wants to evaluate different hypotheses than originally (e.g., $H_{A_5}\not\cap~H_{A_0}$). 

Understanding how users acquire new insights, or develop new analysis questions, has largely been studied with open-ended protocols qualitatively\cite{north2006toward}. 
With a grammar to formally express $H_A$, we can now examine and enumerate the relationships between $H_A$, $H_D$, and $H_V$. For example, does $H_A$ change when certain visualizations are recommended\cite{2017-voyager2}, or with in-situ data augmentation\cite{cashman2020cava}? We discuss these questions as visualization research opportunities in Section~\ref{sec:opportunities}.

\subsection{Formalizing Analysis Tasks in Visualization}
\label{sec:comparison-tasks}

The ability to formalize tasks can provide practical benefits to visualization design and automation\cite{fisher2017making, dimara2021critical}. 
Towards this goal, researchers have proposed a number of task taxonomies\cite{brehmer2013multi, schulz2013design, dimara2018task, kim2018assessing, kerracher2015task}.
While taxonomies are indeed useful for grouping tasks into logical categories, they do not provide formal definitions for the tasks\cite{miller1967task}.
For example, the \textit{Characterize Distribution} task has appeared in a number of low-level task taxonomies\cite{isenberg2013systematic, amar2005low, yi2007toward, chi1998operator, casner1991task}, and while semantically informative, describing a user's task as ``characterize distribution'' does not provide a definition, criteria, or characterization for the distribution.

We demonstrate that our grammar enables the formalization of analysis tasks by providing explicit definitions.
We also illustrate that, in some cases, the process of the formalization can reveal ambiguities in a task or taxonomy.
For example, we can define the task ``characterize distribution'' using our grammar as:
\begin{lstlisting}
    hyp :- func (attr) < theta
   func :- fit_Gaussian | fit_Linear | fit_Beta
\end{lstlisting}

Where the function \lstinline{func} fits data to different distributions (i.e., Gaussian, Linear, Beta) and returns the loss (e.g., mean squared error). The hypothesis is evaluated by checking that the resulting loss is below a threshold (\lstinline{theta}). 
We argue that our grammar to describe ``characterize distribution'' is more precise than current versions in taxonomies, as it provides the specific types of distributions that a user might consider. This grammar also makes explicit the characterization criteria, which is that the loss has to be below some threshold \lstinline{theta}.

Applying our grammar to define other low-level tasks can reveal potential ambiguity in a taxonomy.
For example, the low-level task \textit{Find Anomalies} appears in many visualization task taxonomies, but how to define an anomaly left up to interpretation.
In this case, we are unable to provide a general definition of ``find anomaly'' using our grammar.
Instead, any attempt to define the ``find anomaly'' task results in defining what an anomaly is. For example:
\begin{lstlisting}
    hyp     :- attr1 [id = const1] ^$\not\in$^ [ min, max ])
    min     :- (avg(attr1) - const2 * stdev(attr1))
    max     :- (avg(attr1) + const2 * stdev(attr1))
    const1  :- number
\end{lstlisting}

This grammar defines an anomaly as any value that is outside some number (\lstinline{const2}) of standard deviations away from the average.
Although the exact distance is not specified (\lstinline{const2}), the grammar defines that anomalies are distance-based and the distance is determined using class average and standard deviation. 
We propose that the use of the grammar is more useful than current approaches, as it provides more specificity than the categorical name ``find anomalies.'' 
However, as there is no generalizable way to describe the task, we find that the task itself is ambiguous. We provide a full illustration of how our grammar maps to low-level analytic tasks in the supplemental material.

\section{Demonstration of the Grammar}
\label{sec:demonstration}

To show a practical implementation of our proposed grammar, we use the 2017 VAST Challenge to illustrate the process of expressing sets of analytic questions as a hypothesis space. Finally, we discuss the learning outcomes of this demonstration.

\subsection{2017 VAST Challenge MC1}
\label{sec:use-case-vast}
The VAST Challenges have a rich history in the visual analytics community\cite{cook2014introVAST}. Each challenge is designed to simulate complex, open-ended problems that can be solved through careful data and visual analysis. For this case study, we examine the first mini-challenge (MC1) of the 2017 VAST Challenge. For brevity, we shorten the challenge description and data attribute values in our text. The full description of the challenge can be found online\footnote{\url{https://www.vacommunity.org/VAST+Challenge+2017}}.

\smallbreak 

\noindent 
\textbf{Problem statement:} The population of the Rose-Crested Blue Pipit (a local bird species) is decreasing within a nature preserve. Provide hypotheses to explain any potentially odd behaviors in vehicle activity that might be causing this decrease. 

\smallbreak 

\noindent 
\textbf{Dataset:} The challenge includes two datasets: a map of campsites, ranger stations, gates, etc.~throughout the preserve, and a sensor log of the vehicular activity that has the schema: \lstinline{Timestamp:datetime, Car-id:str, Car-type:str, Gate-name:str}. The metadata as well as a subset of data are:

\begin{itemize}[leftmargin=*,topsep=1pt, partopsep=1pt,itemsep=1pt,parsep=1pt]
    \item \lstinline{Timestamp}: date and time a sensor reading was taken
    \item \lstinline{Car-id}: unique car ID 
    \item \lstinline{Car-type}: vehicle type (e.g., car, ranger) belonging to a given car-id
    \item \lstinline{Gate-name}: name of the sensor (e.g., gate) taking the reading.     Gates can be entrances, general gates, ranger areas, or camping sites
\end{itemize}

\vspace{-0.9em}

\noindent 
{\scriptsize
\begin{center}
\begin{tabular}{l l l l}
    \texttt{Timestamp} & \texttt{Car-id} & \texttt{Car-type} & \texttt{Gate-name}  \\
    5/1/2015 7:50 &	523	& 2-axle-car & general-gate2  \\
    5/1/2015 7:52 &	669	& ranger-truck & ranger-base  \\
    5/1/2015 7:53 &	647	& 2-axle-ranger-truck & entrance4  \\
    5/1/2015 7:58 &	751	& 3-axle-truck	& ranger-stop0 \\
    5/1/2015 7:59 &	523	& 2-axle-truck	& camping1 \\[3pt]
\end{tabular}
\end{center}
}

\subsubsection{Expressing an Analysis Hypothesis Space}
\label{sec:use-case-ahs-mc1}

To express an analysis space for the challenge, we first decompose a set of analysis questions and represent them as a hypothesis grammar. We synthesized these questions based on official Challenge submissions found online\cite{benson:2017:mystery, Buchmuller:2017:vast, cappers:2017:vast, yifei:2017:vast}. 

For simplicity, our grammar assumes that data transformations (joining tables, extraction, cleaning, etc.) have already been applied.
Some of the following examples require deriving new data (e.g., computing \texttt{duration} from two consecutive timestamps) which can be informed through the construction of the grammar. We separately describe the transformations needed for those hypotheses. 

\medbreak 

\newcounter{counter}
\setcounter{counter}{1} 

\noindent 
\textbf{Q\thecounter:} Are park visitors always exiting the park after they enter? 
\begin{lstlisting}[escapeinside={$}{$}]
    hyp :- (mod(count(), 2) = 1)[pred]
   pred :- [Gate-name~"^entrance" & Car-id=const1]
 const1 :- number
\end{lstlisting}
This grammar lets us evaluate whether some park visitor are not passing through an exit gate once entering. 
To do so, the grammar counts the number of check-ins at any entrance gate and checks that it is odd (based on the count modulo 2).   `\verb|~|' tests Gate-name against the regular expression for strings that start (\verb|^|) with ``entrance''.   The non-terminal \texttt{const1} expresses the space of all hypotheses where a unique Car-id passed the entrance gates an odd number of times.




\medbreak 
\stepcounter{counter}
\noindent 
\textbf{Q\thecounter}: Is there illegal activity in the park? 
\begin{lstlisting}[escapeinside={$}{$}]
    hyp :- (expr1 > 0)[!(Car-type~"ranger")]
  expr1 :- count()[Gate-name~"^ranger"]
\end{lstlisting}
This grammar evaluates if non-ranger cars are entering ranger areas, that is, whether vehicles are (illegally) checking in through ranger gates.

\medbreak 
\stepcounter{counter}
\noindent 
\textbf{Q\thecounter}: Is the park more popular during Pipit mating season? 
\begin{lstlisting}
     hyp :- count()[Season="Spring"] > 
            count()[Season=const1]
  const1 :- "Winter" | "Fall" | "Summer"
\end{lstlisting}
This grammar evaluates whether there are more vehicles visiting the park in the Spring than in other seasons. \texttt{Season:string} can be computed based on the month in \texttt{Timestamp}.

\medbreak 
\stepcounter{counter}
\noindent 
\textbf{Q\thecounter}: Are cars speeding when rangers are off duty? 
\begin{lstlisting}
    hyp :- avg(Speed)[Hour^$\in$^[0,7)] > avg(Speed)[Hour^$\not\in$^[0,7)]
\end{lstlisting}
This grammar evaluates whether the average car speed in the park is highest between midnight and 7am. To do so, we first pre-compute \texttt{Speed:float} as the distance between two consecutive gate check-ins of the same car, then divide by the time difference.

\medbreak 
\stepcounter{counter}
\noindent 
\textbf{Q\thecounter}: Are there any outliers in vehicle nightly activity? 
\begin{lstlisting}
   hyp :- KL(DistrEvening, DistrDay) < 10
\end{lstlisting}
This grammar evaluates whether the distribution of gate check-ins at night follows the same distribution during the day. To evaluate these hypotheses, we must first pre-compute a distribution of the number of gate check-ins per hour in the evenings (\texttt{DistrEvening}) and day time (\texttt{DistrDay}).
It then uses KL divergence to confirm that the difference between the two distributions is less than a threshold. 

\medbreak 

\stepcounter{counter}
\noindent 
\textbf{Q\thecounter}: Is the length of visitor stay affecting nesting? 
\begin{lstlisting}
  hyp :- Corr(StayLengths, NestingAmount) < -0.75
\end{lstlisting}
This grammar evaluates whether the duration of a visitor's stay in the park is inversely correlated with nesting activity. We need to pre-compute the average visitor stay length per week or month, and the average amount of nesting over the same time interval.
Although the average visitor stay is computable, there is \textbf{Insufficient Data} (H$_A \not\subset $H$_D$) to compute nesting activity, and thus we cannot verify the hypothesis.

\medbreak 
\noindent 
\textbf{Complete Analysis Hypothesis Space.}
Ignoring H\thecounter, the resultant analysis hypothesis space H$_A$ expresses the above hypotheses, as well as others relevant for addressing the challenge's problem statement:
\begin{lstlisting}
  hyp   :- expr op expr  ([ pred ]) (& hyp)?
  expr  :- func ((expr (, expr)?)?  ) | var
  var   :- attr [pred] | const
  pred  :- attr op const 
  func  :- ! | % | avg | count | Corr | KL 
  op    :- = | < | > | != | ^$\in$^ | ^$\not\in$^
  attr  :- Car-id | Car-type | Gate-name | Distance | 
           Duration | Speed | Time | Hour | Check-in | 
           Check-out | Season | Restricted-area | 
           DistrDay | DistrEvening | StayLengths
  const :- number | str | datetime | boolean
\end{lstlisting}

\subsubsection{Takeaways}
One of the values of using a grammar to express an analysis hypothesis space is that it informs us of whether the current data can appropriately evaluate all statements (i.e. the analysis hypothesis space intersects fully with the data hypothesis space). In \textbf{Q6}, we saw that the hypothesis grammar generates statements that cannot be verified, and therefore has inadequate overlap with the challenge's data hypothesis space.



Furthermore, it is interesting to note that the original data's hypothesis space H$_D$ has little overlap with the analysis hypothesis space H$_A$, as is evident by the numerous \texttt{attr} values not present in the original data. 
This is in line with the motivation of the VAST Challenge: simple analyses with the raw data will not solve the problem statement. It is through the formation of hypotheses, and the respective data transformations, that makes it possible to solve the challenge.

By expressing the first mini-challenge from the 2017 VAST Challenge as a grammar, it becomes clear what data transformations are necessary so that H$_A \subset $ H$_D$. We believe this procedure of forming an analysis hypothesis space, then curating the data appropriately, can aid in traditional exploratory pactices.
Similarly, \textbf{Q6} makes clear that additional data needs to be collected before it can be answered, or in the worst case, that the question \textit{cannot} be answered for the challenge..

In summary, by ``pre-registering'' an analysis hypothesis space, an analyst can know ahead of time (1) the data transformations or feature engineering she will need to perform, and (2) what visualizations can potentially be used to visually evaluate her hypotheses.  In addition to informing whether her analysis space overlaps with the data space, H$_A$ can also act as a proxy for an initial ``design requirements gathering'' before designing a visualization system.

\begin{figure} 
    \centering
    \subfloat[Overlaps between $H_D$ and $H_V$.]{{\includegraphics[width=0.17\textwidth]{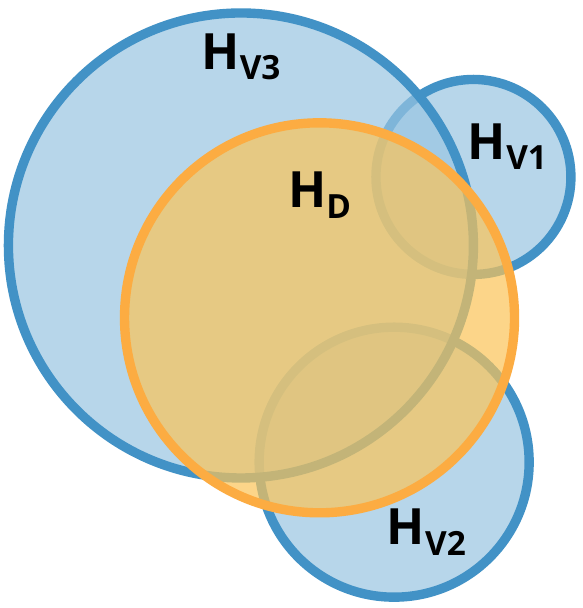} }
    \label{fig:rec_h_d}
    }
    \quad
    \subfloat[Overlaps between $H_A$ and $H_V$.]{{\includegraphics[width=0.2\textwidth]{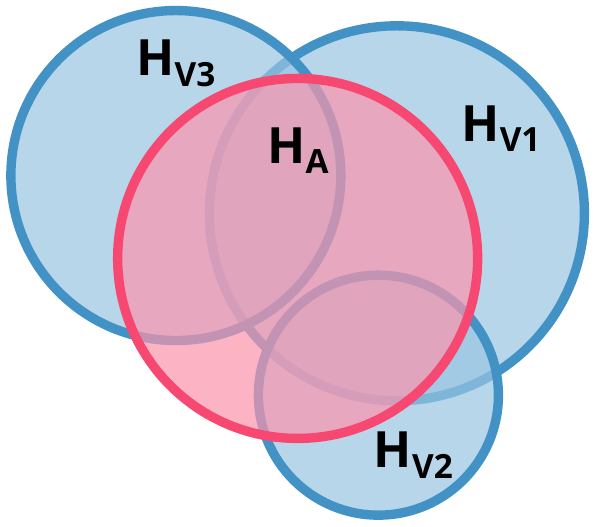} }
    \label{fig:rec_h_a}
    }
    \caption{Using the overlap of hypothesis spaces (Section~\ref{sec:spaces}) to inform visualization recommendations. We describe in detail in Section~\ref{sec:opportunities-rec}.}
\label{fig:hypo-rec}
\end{figure}

\section{Research Opportunities}
\label{sec:opportunities}
Along with the implications of our hypothesis grammar presented in Section~\ref{sec:implications}, we envision multiple research opportunities that can incorporate our grammar such as expressing analyses, recommending and evaluating visualizations, as well as supporting refinement.

\subsection{Hypothesis-Based Visualization Recommendation}
\label{sec:opportunities-rec}

Related to the work proposed by Mackinlay\cite{mackinlay1986automating}, a grammar for tasks can automate visualization recommendation. Research has looked into task-based recommendations of simple visualizations and simple analytic tasks. For example, Saket et al. trained Kopol, a visualization recommender that suggests visualizations to a user based on a given task and data type\cite{saket2018task}. Similarly, Shen et al. proposed a task-based visualization recommendation system\cite{shen2021taskvis}, which maps a collection of analytic tasks (e.g., `correlate,' `find extrema,' `cluster') to generate and rank candidate visualizations, then outputs the results to a user. 

A benefit of these recommenders is the reduced effort and expertise required of users to design task-appropriate visualizations. If a user's analysis task is simple enough, we are now able to provide qualified visualizations to satisfy those tasks. This is a significant step forward from custom, manual visualization creation.

Complementing these efforts, we believe the opportunity to recommend visualizations based on $H_A$ and $H_D$ can be equally beneficial.
To illustrate this, consider Figure~\ref{fig:hypo-rec}. Given a user's dataset and the hypotheses it can express ($H_D$), visualizations can be recommended such that their spaces overlap the most appropriately with $H_D$ (e.g., $H_V3 > H_V2 > H_V1$), as seen in Figure~\ref{fig:rec_h_d}. Similarly, visualizations can be recommended based on which of their hypotheses spaces $H_V$ overlaps the most with $H_A$, which may result in different recommendations (e.g., ($H_V1 > H_V3 > H_V2$), shown in Figure~\ref{fig:rec_h_a}.  
%
 %
 
This opens new directions in ongoing efforts to understand ``task-appropriate" visualizations. Using our proposed grammar-based approach, researchers can begin to study how to automatically express a user's analysis goals ($H_A$), a candidate visualization's analysis capabilities ($H_V$), the data ($H_D$) best mapped to visualizations, and how to use their intersections to inform recommendations. 

\hfill



\subsection{Visualization Evaluation}
Similar to the topic of hypothesis-based recommendations for visualizations is the opportunity to formally evaluate visualizations on their ability to support a grammar of hypotheses. We observe two directions that this research opportunity can take.

First, we are now able to evaluate a visualization's efficacy in both expressing and verifying hypotheses. As we showed in Section~\ref{sec:vis-space}, a visual encoding or visualization (e.g., a bar chart) is capable of expressing a set of hypotheses (\lstinline{bar1 > bar2}, etc.) -- these hypotheses span the visualization's hypothesis space. Just as some visualizations are better suited for visualizing certain data types\cite{mackinlay1986automating}, or supporting certain simple analytic tasks\cite{saket2018task}, there will be visualizations better suited for expressing and verifying certain hypotheses. For example, to verify hypotheses that pose part-to-whole relationships, a pie chart will be preferable to a line chart. 
Experiments to quantify efficacy of expressing and verifying hypotheses across different visualizations is a rich and open area for future research.   


Second, we can evaluate a visualization system by its ability to help users navigate and refine their hypothesis space ($H_A$). Theoretical models for how users learn from visualization tools, e.g., through knowledge generation\cite{sacha2014knowledge}, the data-frame theory\cite{klein2007data}, or sensemaking\cite{pirolli2005sensemaking} have spurred new frameworks for testing the efficacy of interactive visualization tools. In particular, in terms of these tools' abilities to help users generate and validate knowledge, as well as fit data into a frame (and vice versa), and make sense of data, respectively.
Similar in spirit to these models, by tracking how an analysis hypothesis space changes through the use of a tool (e.g., via exploration, as discussed in Section~\ref{sec:demonstration-partial}), we can similarly understand how systems support users through searching, refining, and expanding hypothesis spaces.

\subsection{Guiding Users in Analysis with Grammars \& NLP}
\label{sec:opportunities-refinement}

Another area in which our grammar presents research opportunities is guiding users in analysis. Specifically, helping users refine broad, vague analysis goals into executable hypotheses. 
Users often begin an analysis session with a vague notion of their overarching analysis goal or task. In these cases, modeling analysis tasks as hypotheses allows us to asses and refine under-developed hypotheses into well-defined, testable hypotheses. 

\begin{table}[t]\centering
\small
\renewcommand{\arraystretch}{1.1}
\sffamily
\resizebox{.9\linewidth}{!}{%
\begin{tabular}{p{0.05\linewidth} p{0.8\linewidth}}
        \toprule
        \textbf{Level} & \textbf{Criterion} \\ 
        \midrule
        0 & No explanation provided. A nonsense statement, question, observation, or single inference about a single event, person, or object. \\
        1 & An inappropriate (e.g., irrational, not falsifiable) explanation. \\
        2 & Partial appropriate explanation (e.g., incomplete). \\
        3 & Appropriate explanation relating at least two variables in general or nonspecific terms. \\
        4 & Precise explanation with qualification of the variables. A specific relationship is provided. \textbf{This is a hypothesis}. \\
        5 & In addition to level 4, a test is given. \\
        6 & An explicit statement of a (precise) test of a hypothesis is given.\\
        \bottomrule
    \end{tabular}}
    \caption{Sunal and Haas' \textit{Hypothesis Quality Scale}\cite{sunal2002social} for writing and evaluating students' hypotheses. 
    }
\label{tab:hyp_quality_scale}
\end{table}

One method to do so is through the use of \textit{Hypothesis Quality Scales}, such as Sunal and Haas'\cite{sunal2002social}, shown in Table~\ref{tab:hyp_quality_scale}. This quality scale is used to provide rankings for a student's hypothesis such that they can be refined and improved upon.   
With a hypothesis quality scale, vague or under-specified analysis tasks can be thought of as underdeveloped scientific hypotheses. Scales such as Sunal and Haas' provide the mechanism for adding clarity to hypotheses and analysis tasks.

Another possible approach to refining analysis goals is by combining our hypothesis grammar with advancements in NLP and natural language interfaces (NLIs)\cite{Aurisano2016Articulate2, Narechania2020NL4DVAT, mitra2022facilitating}. These interfaces can help identify areas of ambiguity in a hypothesis that are apt for refinement.  
For example, in FlowNL\cite{huang2022flownl}, users can query a flow visualization system in plain English to generate relevant graphical views (e.g., ``\textit{where are hurricanes occurring?}''). When their natural language parser does not recognize the user's query, the system will ask the user to clarify a particular word or concept (``\textit{what does `hurricane' represent in the data?}'') -- then stores the user's definition to be reused throughout the user's analysis session. Similar work could be done with the integration of a hypothesis grammar. For example, if the user wants to pose partially-specified hypotheses, an NLI could guide users to refine their analysis questions, goals, and/or hypothesis.

\section{From Exploratory to Hypothesis-Driven Analysis}
\label{sec:comparison}
The work presented in this paper focuses on how a grammar-based approach for hypotheses can be used to operationalize analysis tasks. 
This approach offers the benefit of contributing towards the growing body of work in belief elicitation\cite{mahajan2022vibe} and ``hypothesis-first'' analysis\cite{cockburn2018HARK, kim2019bayes}.
In visual belief elicitation, a user is asked to first convey their beliefs, assumptions, or predictions about data shown on a visualization.  
A popular example is the New York Times' ``You Draw It'' series\cite{aisch2015you} -- users are given a blank (or partially complete) line chart, and prompted to draw their prediction for a given question, e.g., ``\textit{how does a family's income predict their children's college chances?}'' 


Empirical research into the effects of visual belief elicitation\cite{koonchanok2023visual} and ``hypothesis-first'' analysis shows promise. 
Users are more engaged with visualizations\cite{hohman2020communicating}, conduct more deliberate analyses\cite{prophecy2021koonchanok}, discover fewer spurious and false discoveries\cite{zhao2017controlling, zgraggen2018investigating} than with exploratory data analysis\cite{tukey1977exploratory}, and have more accurate recall of data\cite{kim2017explaining, hullman2018imagining}.
Consequently, a multitude of conceptual frameworks, cognitive models, and design spaces have been proposed to describe the processes of hypothesis- and belief-first analysis\cite{choi2019concept, kim2019bayes, mahajan2022vibe}. 

Hullman and Gelman\cite{designing2021hullman} emphasize the need for theoretical models to unite and complement traditional exploratory practices\cite{tukey1977exploratory, keim2006challenges} and visualization tools\cite{2016-voyager, 2017-voyager2}. A hypothesis-driven model for visual analytics can contribute towards addressing this gap. As Hullman and Gelman similarly advocate, we believe hypothesis-first interfaces could mitigate analysis risks, aid in interpreting uncertainty, encourage the ability to infer correct conclusions about data, and 
incorporate the many empirical findings discussed above\cite{zgraggen2018investigating, koonchanok2023visual, prophecy2021koonchanok}.
In terms of advancements to visualization theory, a hypothesis-driven approach (which we will refer to as \textbf{HDA}) would be able to support new considerations for the processes of knowledge generation and verification\cite{sacha2014knowledge, kell2004here}.

\begin{wrapfigure}{l}{.24\textwidth}
    \centering
    \vspace{-7pt}
    \includegraphics[width=.25\textwidth]{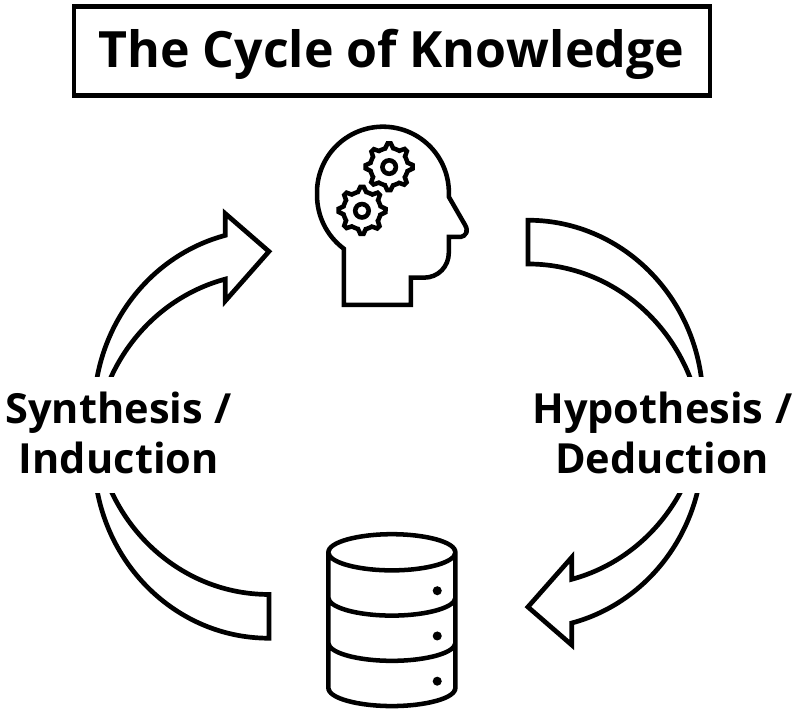} 
    \vspace{-21pt}
\end{wrapfigure}
The figure to the left illustrates how the scientific community considers the cycle of knowledge\cite{kell2004here}. 
Commonly, visualization and data analysis practices follow the principle of exploratory data analysis (EDA)\cite{tukey1977exploratory}, or a ``data-first'' approach, where an analyst begins analysis by first analyzing the data (illustrated as \textit{Synthesis/Induction} in the figure).
An HDA path would work in reverse, promoting a ``hypothesis-first'' principle\cite{choi2019concept, designing2021hullman, kim2019bayesian}, where analysis begins with an analyst posing \textit{a priori} hypotheses -- similar to the scientific method (illustrated as \textit{Hypothesis/Deduction} in the figure). We posit that these processes should work in tandem:
a data-first approach should result in meaningful hypotheses and insights, and a hypothesis-first approach should result in hypotheses that are validated with data.

Given the multitude of observed benefits for `hypothesis-first' analysis\cite{koonchanok2023visual, hohman2020communicating, prophecy2021koonchanok, zhao2017controlling, zgraggen2018investigating, kim2017explaining, hullman2018imagining, choi2019concept, mahajan2022vibe}, one may wonder why HDA is not an established, fundamental methodology to visual analytics. Arguments have been made that users do not (and should not) necessarily start analysis with any \textit{a priori} hypotheses formed\cite{yanai2020hypothesis}. 
However, we find ourselves in agreement with Felin et al.\cite{felin2021data}, among others\cite{miller2003presumptions, designing2021hullman, kim2019bayesian}, who argue that ``\textit{there is no such thing as hypothesis-free data exploration,}'' and that ``\textit{even informal hunches or conjectures are types of proto-hypothesis}.'' 

Another hurdle to HDA in visual analytics is the seemingly difficult integration of a user's complex beliefs for data. 
Currently, there is no operational mechanism to implement well-defined \textit{and} partially-specified hypotheses. Despite the conceptual frameworks and design spaces\cite{choi2019concept, kim2019bayes, mahajan2022vibe} that have been contributed, a practical gap still exists in the \textit{formalization} of ``hypothesis-first'' visual analytic systems and workflows\cite{designing2021hullman}.
We posit that our grammar for hypotheses (as well as its ability to support `unknowns', Section~\ref{sec:implications}) can provide a baseline for tools to seamlessly integrate a user's hypotheses. 
Unlike previous research, where hypothesis-first analysis is conducted and studied in ``Wizard of Oz'' experiments (e.g.,\cite{choi2019visual, koonchanok2023visual}), we offer an operationalizable method to realize these processes. 

For now, the exact relationship and boundaries between EDA and HDA are unclear. 
Perhaps hypothesis-driven analysis should be a concrete \textit{component} of EDA, formalizing a boundary in which users step from ``exploratory'' to ``confirmatory'' analysis phases\cite{tukey1977exploratory}. 
In this representation, HDA would provide a coupling of the processes for EDA and CDA. Regardless, HDA offers a realistic way to conceptualize the `hypothesis-first' analysis process, and our grammar for hypotheses offers a concrete step towards operationalizing systems supportive of it.

\section{Limitations}
\label{sec:discussion}

In this paper, we advocate for operationalizing analysis tasks to disambiguate their use in visualization. We pose one possible grammar-based solution using the concept of \textit{scientific hypotheses}.
Below, we provide considerations of the limitations and unknowns for our proposed hypothesis grammar to make tasks operational.


\subsection{Is this Grammar Enough?}
A formal, universal definition of ``task" currently does not exist in visualization literature. Numerous domain-specific tasks have been proposed (e.g.,~\cite{kerracher2015task, miksch2014matter, laha2015classification, etemadpour2015user, sarikaya2018scatterplots}), as have tasks specific to particular applications or usage contexts (e.g.,~\cite{isenberg2013systematic, sedlmair2012design, chang2010learning, dimara2018task}). 
Consequently, the concept of ``tasks'' in visualization and visual data analysis spans a considerable number of actions, procedures, and needs.

A single hypothesis grammar cannot possibly operationalize \textit{all} definitions of ``tasks'' in visualization. Our grammar is constrained to expressing analysis questions (as testable \textit{relationships} between data variables) in the space of (tabular) data, and (mostly static) visual encodings. For example, our grammar can not encode tasks like ``\textit{Enjoy}\cite{brehmer2013multi},'' as these tasks are not analytical questions in nature. 

With this said, we believe \textbf{a grammar-based approach to make analysis tasks operational is paramount}. Whether that approach is using our hypothesis grammar or a future `task-based' grammar, it will provide new methods to push visualization design and theory forward using a new shared language. We envision many grammar-based approaches for tasks in the future, and hope that our hypothesis grammar provides a baseline to improve and reflect upon. 



\noindent 
\subsection{Boundedness of the grammar}
Our grammar is not capable of expressing \textit{every} possible hypothesis that could be formulated with the English language. Hypotheses are a complex subset of natural language and have been an ongoing object of study. Previous grammars for hypotheses have been contributed in literature\cite{chomsky:1956:three, van2007issues}, but are typically limited to a specific domain application (e.g., electrical circuits\cite{kroeze2019automated}). Similarly, our hypothesis grammar is restricted to performing analysis and posing questions and statements about data.


\noindent 
\subsection{Assumptions for the data}
We assume the data is relational and can be accessed from a database (or a table). We also assume the data has been transformed, cleaned, and is not missing any values required by the grammar. This is obviously not always the case with ``real-world'' analysis. 
The existence of data transformation languages such as SQL and non-relational query languages suggest that extending the grammar to support data pre-processing is possible.  Resulting trade-offs between grammar simplicity and expressiveness must be carefully balanced.   


\noindent 
\subsection{Qualifying human judgement}
Our hypothesis grammar does not take into account uncertain criteria for human judgement. 
For example, the statement ``\textit{action movies are better than comedies}'' could possibly be formed as a hypothesis, but it is subjective. Our grammar does not take these hypotheses into consideration but instead focuses on quantifiable \lstinline{true} or \lstinline{false} statements that can be posed about data, driven by the notion of \textit{scientific hypotheses}. 


\noindent 
\subsection{Visualization's ability to verify hypotheses}
We postulate that visualizations can confirm or reject some subset of hypotheses given {$H_{V}$}, as illustrated by Wickham et al.~with their \textit{LineUp} technique\cite{graphical2010wickham}. However, currently we cannot enumerate \textit{all} hypotheses that a visualization can be used to answer. There are many open-ended questions that our grammar therefore introduces. If a single visualization is capable of verifying a particular hypothesis, what is the translation to visualization tools in which many visualizations can be shown or suggested to users? Are there some visualizations that are more indicative of supporting a hypothesis than others? Prior work has sought to understand the effectiveness of a visualization in supporting \textit{tasks}\cite{saket2018task}. In future work we aim to investigate if the same can be done for hypothesis spaces.

%
\noindent 
\subsection{Extensions to other concepts in visual analysis}
In this work, we demonstrate applications of our grammar with real-world analytic challenges. However, there are other important concepts of data analysis and visualization that we did not have space to address. For example, how uncertainty affects data understanding\cite{song2018s}, data augmentation\cite{cashman2020cava}, perceptual models and experiments\cite{dosher2017visual}, and so on. We hope that our formalism for hypotheses in visualization can provide the springboard for research in all of these areas.

\section{Conclusion}
\label{sec:conclusion}

This paper presented a grammar-based approach for operationalizing analysis tasks in visualization. The grammar draws from the science and education literature, and provides a formal language to express (scientific) hypotheses over datasets.   We showed how sets of hypotheses (parsable by our grammar) make up a {\it hypothesis space}, and can be used to represent the hypotheses a specific dataset can evaluate ({data hypothesis space}), the hypotheses a visualization can evaluate ({visualization hypothesis space}), and the hypotheses a user wants to evaluate through analysis ({analysis hypothesis space}).

We illustrated how the use of our grammar can unify a user's analysis goals, the available data, and the capabilities of a visualization system. 
We demonstrated how 
it can allow partial specification, and how it can formalize analysis tasks commonly used in visualization research. We provided a case study on the use of our grammar with the 2017 VAST Challenge, 
which helped 
identify necessary data transformations, as well as questions that were unanswerable by the data. 
Finally, we discussed the practical benefits of our hypothesis grammar to aid in future research directions, including a new hypothesis-driven approach (HDA) for visual analytics. 
Overall, we believe a grammar to articulate tasks will promote better analysis clarity, help evaluate visualization systems, and support hypothesis-based reasoning in visual analytics.



\bibliographystyle{vis/abbrv-doi}
\bibliography{references}

\end{document}